# A CRYPTOGRAPHIC IMAGE ENCRYPTION TECHNIQUE FOR FACIAL-BLURRING OF IMAGES

Quist-Aphetsi Kester, MIEEE, Lecturer Faculty of Informatics, Ghana Technology University College, PMB 100 Accra North, Ghana
Phone Contact +233 209822141 Email: kquist-aphetsi@gtuc.edu.gh / kquist@ieee.org

## Abstract

Protection of faces in pictures and videos of people in connection with sensitive information, activism, abused cases and others on public broadcasting media and social networks is very important. On social networks like YouTube, Facebook, Twitter and others, videos are being posted with blurring techniques of which some of them cannot be recoverable. Most blurring techniques used can easily be recoverable using off-the-shelf software. The ones that are difficult to be recovered also can easily be used by abusers and other wrong doers.

This paper proposes an image encryption technique that will make it possible for selected facial area to be encrypted based on RGB pixel shuffling of an m*n size image. This will make it difficult for off-the-shelf software to restore the encrypted image and also make it easy for the law enforcement agencies to reconstruct the face back in case the picture or video is related to an abuse case. The implementation of the encryption method will be done using MATLAB. At the end, there will be no change in the total size of the image during encryption and decryption process.

## Introduction

Identity protection is very crucial in today's digital media world where courts, the media and other institutions as well as individuals would like to share information under a protected cover. People who will like to testify against criminal cases in public hearings in which full protection from the public cannot be guaranteed by the state institutions required to be visually protected. Most of the enhancement image processing techniques can be used to recover blurred images making most of the images and video vulnerable to these techniques. Hence there is a need for a more sophisticated approach based on modern day cryptographic techniques in solving these problems.

Modern day cryptography entails complex and advance mathematical algorithms that are applied to encryption of text and images as well as other file formats. Cryptographic techniques for image encryption are normally based on the RGB pixel displacement where pixels of images are shuffled to obtain a cipher image [1].

Encryption of messages in this modern age of technology becomes necessary for ensuring that data sent via communications channels become protected and made difficult for deciphering. Enormous number of transfer of data and information takes place through internet, which is considered to be most efficient though it's definitely a public access medium. Therefore to counterpart this weakness, many researchers have come up with efficient algorithms to encrypt this information [2].

This paper proposes an image encryption technique that will make possible for selected facial area to be encrypted based on RGB pixel shuffling of an m*n size image. At the end, there will be no change in the total size of the image during encryption and decryption process.

The paper has the following structure: section II consist of related works, section III gives information on the methodology employed for the encryption and the decryption process, section IV presents the algorithms employed to come out with a cipher for the encryption process, section V gives explains the algorithm mathematically by showing the step by step manipulation and shuffling of the image pixels, section VI provided the architectural summary of the encryption process using flow charts, section VII consist of the simulated results as well as graphical analysis and section VIII concluded the paper.

## Related Works

A new cryptographic scheme proposed for securing color image based on visual cryptography scheme was done by Krishnan, G.S. and Loganathan, D. A binary image was used as the key input to encrypt and decrypt a color image. The secret color image which needs to be communicated was decomposed into three monochromatic images based on YCbCr color space. Then these monochromatic images were then converted into binary image, and finally the obtained binary images were encrypted using binary key image, called share-1, to obtain the binary cipher images. During their encryption process, exclusive OR operation was used between binary key image and three half-tones of secret color image separately. These binary images were combined to obtain share-2. In the decryption process, the shares were decrypted, and then the recovered binary images were inversed half toned and combined to get secret color image. [3]

With extended Visual Cryptography, which is a method of cryptography that reveals the target image by stacking meaningful images. Christy and Seenivasagam proposed a method that uses Back Propagation Network (BPN) for extended visual cryptography. BPN was used to produce the two shares. The size of the image produced was the same as that of the original image. [4]

A k-out-of-n Extended Visual Cryptography Scheme (EVCS) is a secret sharing scheme which hides a secret image into n shares, which are also some images. The secret image can be recovered if at least k of the shares are superimposed, while nothing can be obtained if less than k shares are known. Previous EVCS schemes are either for black-and-white images or having pixel expansion. Wu, Xiaoyu, Wong, Duncan S. and Li, Qin proposed the first k-out-of-n EVCS for color images with no pixel expansion. The scheme also improved the contrast of the n shares and the reconstructed secret image (i.e. the superimposed image of any k or more shares) by allowing users to specify the level of each primary color (i.e. Red, Green and Blue) in the image shares as well as the reconstructed secret image. [5]

Kester, QA proposed a cryptographic algorithm based on matrix and a shared secrete key.[6]. Which was further applied encryption and decryption of the images based on the RGB pixel [2].

Shujiang Xu,Yinglong Wang , Yucui Guo and Cong Wanga proposed a novel image encryption scheme based on a nonlinear chaotic map (NCM) and only by means of XOR operation. There were two rounds in the proposed image encryption scheme. In each round of the scheme, the pixel gray values were modified from the first pixel to the last pixel firstly, and then the modified image was encrypted from the last pixel to the first pixel in the inverse order. In order to accelerate the encryption speed, every time NCM was iterated, n (n>3) bytes random numbers were used to mask the plain-image. And to enhance the security, a small perturbation was given to the parameters of the NCM based on the last obtained n bytes modified elements before next iteration. [7]

Ruisong Ye and Wei Zhou proposed a chaos-based image encryption scheme where one 3D skew tent map with three control parameters were utilized to generate chaotic orbits applied to scramble the pixel positions while one coupled map lattice was employed to yield random gray value sequences to change the gray values so as to enhance the security. Experimental results have been carried out with detailed analysis to demonstrate that the proposed image encryption scheme possesses large key space to resist brute-force attack and possesses good statistical properties to frustrate statistical analysis attacks. And at the end, the proposed scheme utilizes the 3D skew tent map to shuffle the plain-image efficiently in the pixel Positions permutation process and it employed the coupled map lattice system to change the gray values of the whole image pixels greatly.[8]

With the exceptionally good properties in chaotic systems, such as sensitivity to initial conditions and control parameters, pseudo-randomness and ergodicity, chaos-based image encryption algorithms have been widely studied and developed in recent years. Standard map is chaotic and it can be employed to shuffle the positions of image pixels to get a totally visual difference from the original images.

Ruisong Ye,Huiqing Huang proposed two novel schemes to shuffle digital images. Different from the conventional schemes based on Standard map, they disordered the pixel positions according to the orbits of the Standard map. The proposed shuffling schemes didn't need to discretize the Standard map and own more cipher leys compared with the conventional shuffling scheme based on the discretized Standard map. The shuffling schemes were applied to encrypt image and disarray the host image in watermarking scheme to enhance the robustness against attacks. [9]

Amnesh Goel and Nidhi proposed contrastive methods to encrypt images by introducing a new image encryption method which first rearranges the pixels within image on basis of RGB values and then forward intervening image for encryption. [10]

Image Encryption Based on Explosive Inter Pixel Displacement of the RGB Attribute of a Pixel: In this method focus was more on the inter pixel displacement rather than just manipulation of pixel bits value and shifting of pixel completely from its position to new position. RGB value of pixel was untouched in this method, but R value of pixel jumps to another location horizontally and vertically same as in chaotic method. In the similar manner, G and B values of pixel [11].

Nishiyama, M., Hadid, A.; Takeshima, H., Shotton, J., Kozakaya, T. and Yamaguchi, O worked on Facial Deblur Inference Using Subspace Analysis for Recognition of Blurred Faces where they proposed a novel method for recognizing faces degraded by blur using deblurring of facial images.[12]

With the proposed method in this paper, the shuffling of the image will be done by solely displacing the RGB pixels and also interchanging the RGB pixel values. At the end the total image size before encryption will be the same as the total image size after encryption.

## Methodology

In this method, the facial selected portion of the image used will have their RGB colors extracted from then and then encrypted to have a ciphered image portion. The ciphering of the image for this paper will be done by using the RBG pixel values of the selected portion of the images. There are no changes of the bit values and there is no pixel expansion at the end of the encryption process. Instead the numerical values are transposed, reshaped and concatenated with the RGB values shifted away from its respective positions and the RGB values interchanged in order to obtain the cipher image. This implies that, the total change in the sum of all values in the image is zero.

The image is looked at as a decomposed version in which the three principle component which forms the image is chosen to act upon by the algorithm. The R-G-B components can be considered as the triplet that forms the characteristics of a pixel. The pixel is the smallest element of an image which can be isolated and still contains the characteristic found in the image. The RGB values are shifted out of its native pixel and interchanged within the image boundaries by the algorithmic process. The Shift displacement of the R G and B Values known termed as the component displacement factor array which is different for R, G and B.

With the proposed method in this paper, the shuffling of the image will be ultimately done by solely displacing the RGB pixels and also interchanging the RGB pixel values.
.

## The Algorithm

1. Start
2. Import data from the selected portion and create an image graphics object by interpreting each element in a matrix.
3. Extract the red component as ‘r’
4. Extract the green component as ‘g’
5. Extract the blue component as ‘b’
6. Get the size of r as [c, p]
7. Let r =Transpose of r
8. Let g =Transpose of g
9. Let b =Transpose of b
10. Reshape r into (r, c, p)
11. Reshape g into (g, c, and p)
12. Reshape b into (b, c, and p)
13. Concatenate the arrays r, g, b into the same dimension of ‘r’ or ‘g’ or ‘b’ of the original image.
14. Finally the data will be converted into an image format to get the encrypted image.
15. The inverse of the algorithm will decrypt the encrypted image back into the plain image.

## The Mathematical Explanation

Step1.Start
Step2.Import data from the selected portion of the image and create an image graphics object by interpreting each element in a matrix.
Let Q= an image=Q(R, G, B)
Q is a color image of m x n x 3 array

$$\begin{pmatrix} R & G & B \\ r_{i1} & g_{i2} & b_{i3} \\ \vdots & \vdots & \vdots \\ r_{n1} & g_{n2} & b_{n3} \end{pmatrix}$$

(R, G, B) = m x n
Where R, G, B ∈ I
(R o G) i j = (R) ij .(G) ij
Where R= r_i1 = first value of R
r= [ri1] (i=1, 2… m)
x ∈ r_i1 : [a, b]= {x ∈ I: a ≤ x ≥ b}
a=0 and b=255
R= r= Q(m,n,1)

Where G= g_i2 = first value of G
g= [gi2] (i=1, 2... m)
x ∈ g : [a, b]= {x ∈ I: a ≤ x ≥ b}
a=0 and b=255
G= g= Q(m,n,1)

And B= b_i3 = first value of B
g= [bi3] (i=1, 2... m)
x ∈ b_i1 : [a, b]= {x ∈ I: a ≤ x ≥ b}
a=0 and b=255
B=b= Q (m, n, 1)

Such that R= r= Q (m, n, 1)

Extract the red component as ‘r’

Let size of R be m x n [row, column] = size (R) = R (m x n)

$$rij = r = Q(m, n, 1) = \begin{pmatrix} R \\ r_{i1} \\ \vdots \\ r_{in} \end{pmatrix}$$

Step4. Extract the green component as ‘g’

Let size of G be m x n [row, column] = size (G) = G (m x n)

$$gij = g = Q(m, n, 1) = \begin{pmatrix} G \\ g_{i2} \\ \vdots \\ g_{n2} \end{pmatrix}$$

Step5. Extract the blue component as ‘b’
Let size of B be m x n [row, column] = size (B) = B (m x n)

$$bij = b = Q(m, n, 1) = \begin{pmatrix} B \\ b_{i3} \\ \vdots \\ b_{n3} \end{pmatrix}$$

Step6. Get the size of r as [c , p]

Let size of R be m x n [row, column] = size (r) =r (c x p)

Step7. Let r =Transpose of r

$$r = \begin{pmatrix} R & & & & \\ r_{i1} & \dots & \dots & \dots & r_{n1} \end{pmatrix}$$

Step8. Let g =Transpose of g

$$g = \begin{pmatrix} G & & & & \\ g_{i3} & \dots & \dots & \dots & g_{n3} \end{pmatrix}$$

Step9. Let b =Transpose of b

$$b = \begin{pmatrix} B & & & & \\ b_{i2} & \dots & \dots & \dots & b_{n2} \end{pmatrix}$$

Step10. Reshape r into (r, c, p)

$$r = reshape(r, c, p) = \begin{pmatrix} R \\ r_{i1} \\ \vdots \\ r_{in} \end{pmatrix}$$

Step11. Reshape g into (g ,c ,p)

$$g = reshape(g, c, p) = \begin{pmatrix} R \\ r_{i1} \\ \vdots \\ r_{in} \end{pmatrix}$$

Step12. Reshape b into (b ,c ,p)

$$b = \text{reshape}(b, c, p) = \begin{pmatrix} R \\ r_{i1} \\ \vdots \\ \vdots \\ r_{in} \end{pmatrix}$$

Step13.Concatenation of the arrays r, g, b into the same dimension of 'r' or 'g' or 'b' of the original image

$$= \begin{pmatrix} R & G & B \\ r_{i1} & g_{i2} & b_{i3} \\ \vdots & \vdots & \vdots \\ \vdots & \vdots & \vdots \\ r_{n1} & g_{n2} & b_{n3} \end{pmatrix}$$

Step14.Finally the data will be converted into an image format to get the encrypted image.The inverse of the algorithm will decrypt the encrypted image.

# The architectural summary of the encryption process using flow chart diagram

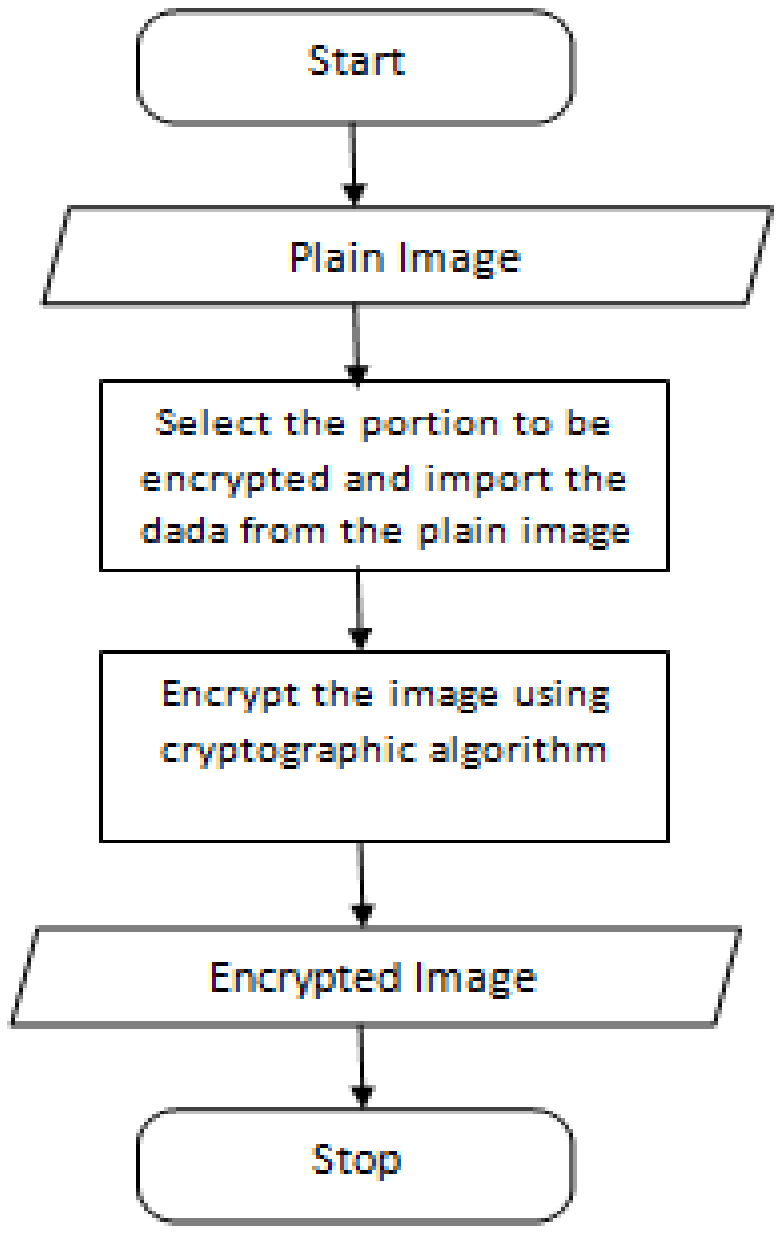

**Figure 1. Flow chart diagram for the encryption process**

# Simulated Results

The simulation of the above algorithm was performed on the MATLAB Version 7.0.0.1. The plain image size used was m x n. The MATLAB code for the algorithm was written and tested the output is shown below.

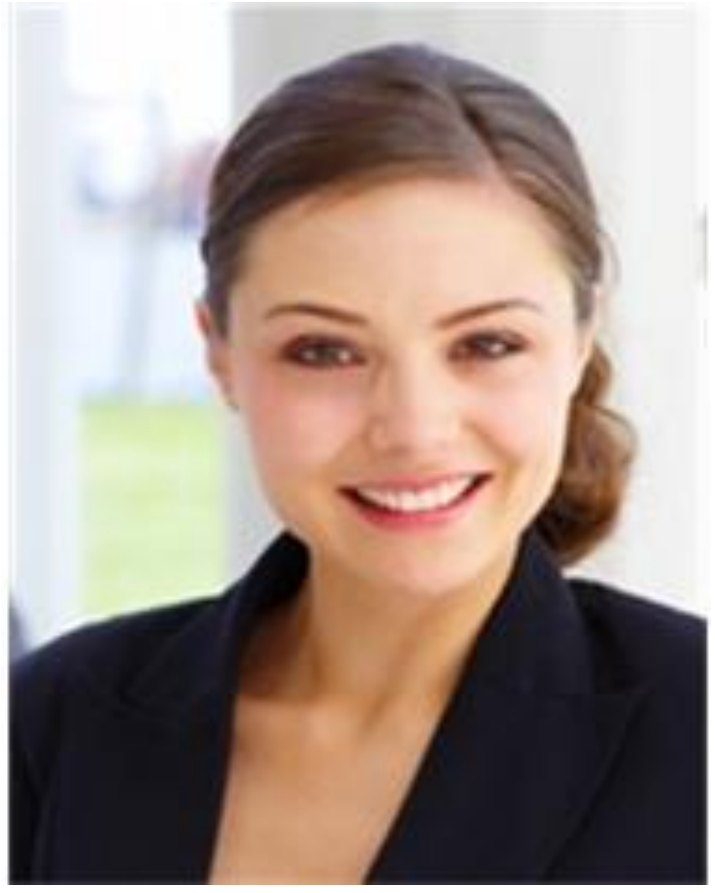

**Figure 2. Plain Image**

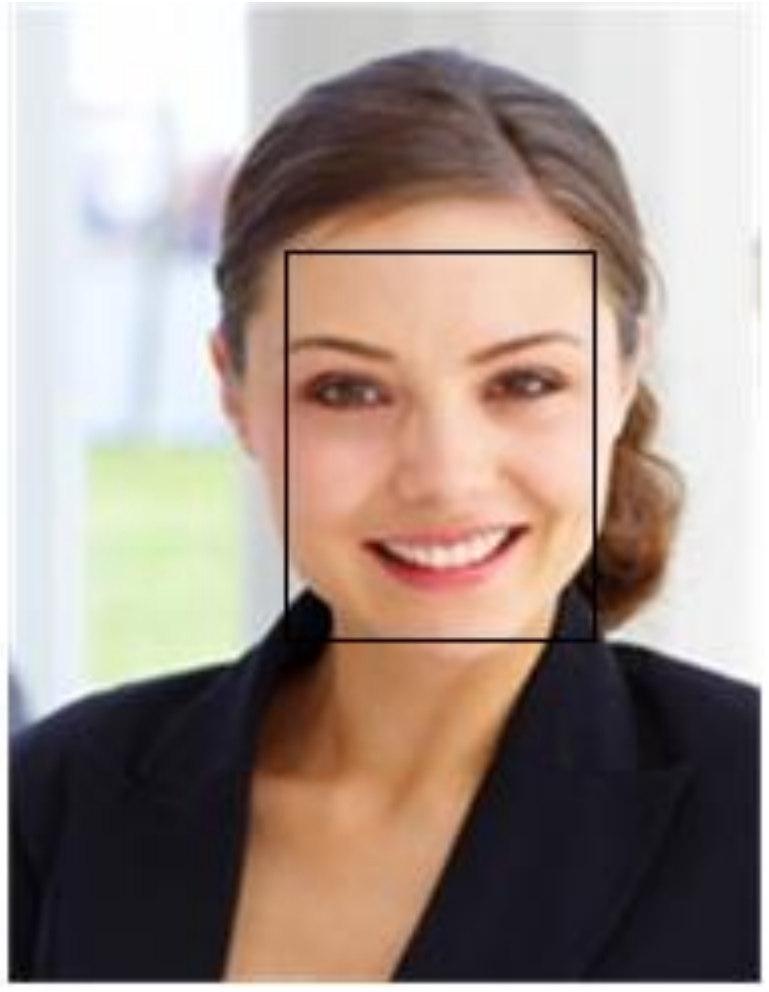

**Figure 3. Selected portion of the image**

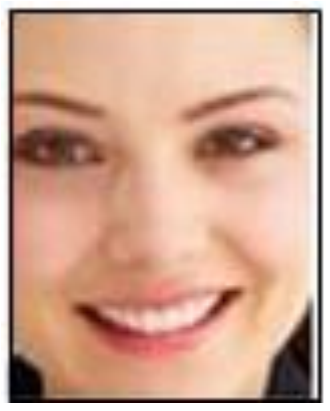

**Figure 4. Selected portion taken out**

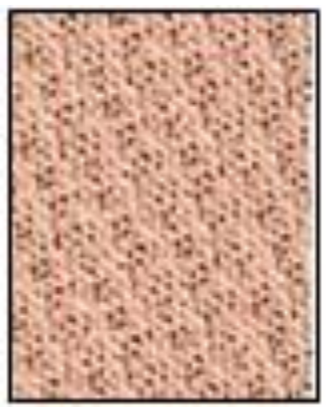

**Figure 5. Encrpted portion of the selected portion**

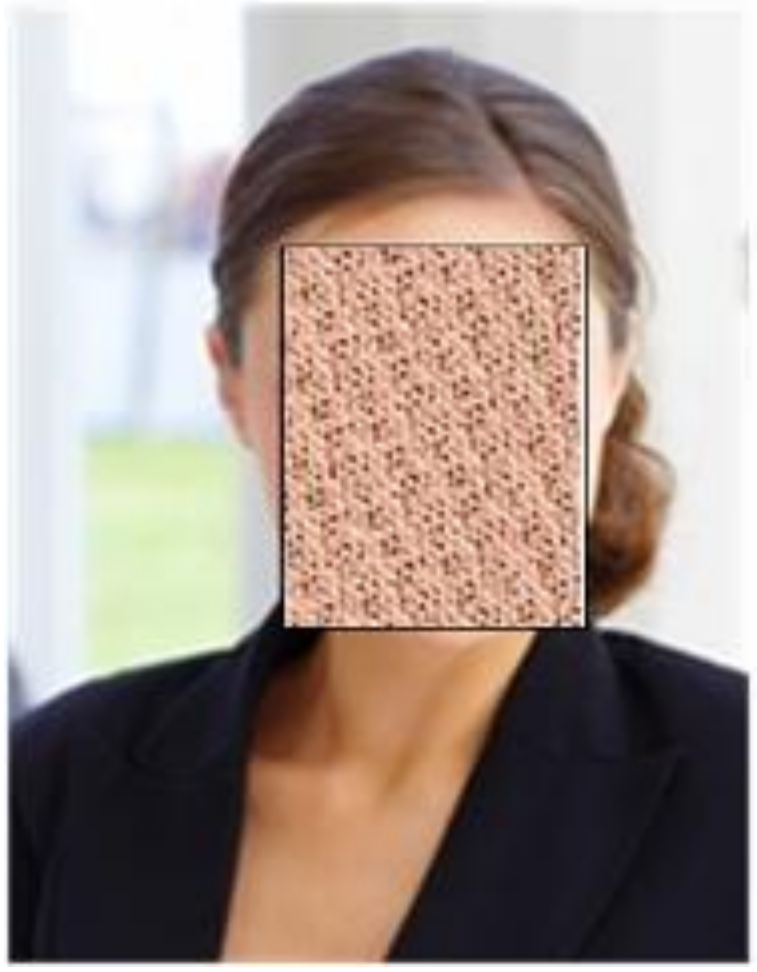

**Figure 6. Final output of the encrypted portion,**

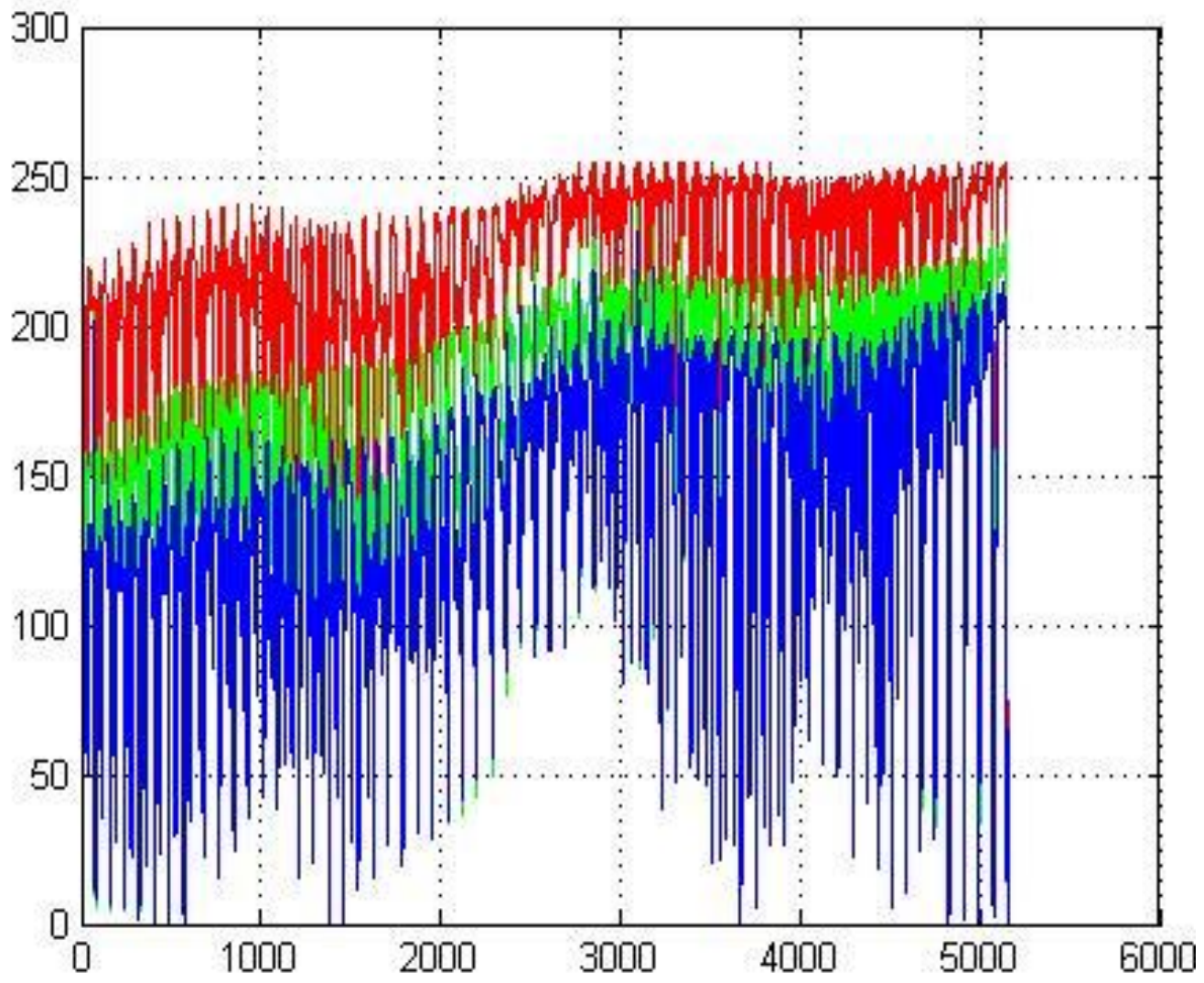

**Figure 7. RGB graph of figure 4**

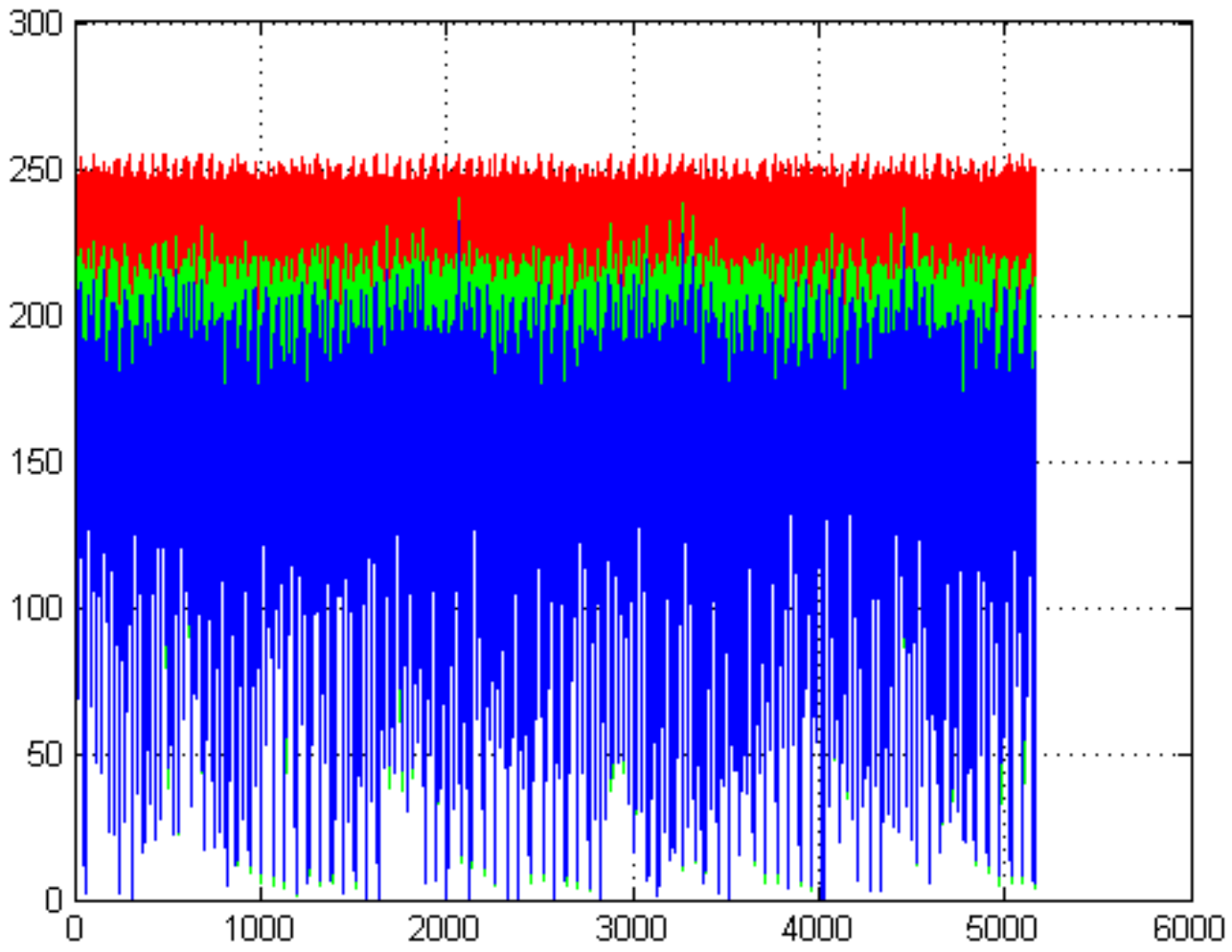

**Figure 8. RGB graph of figure 5**

Figure 2 is the original image to be used. From figure one a dimension of X by Y was selected out of the image.
X=x(x1: x2) and Y=y(y1:1y2)
Which implies the r component r= r(x,y)
This implies r=r(X,Y)
r=r(x(x1: x2), y(y1:1y2))
r=r(X[i=a]:X[i=b],Y[i=c]:Y[i=d])
where i= integer and X[i] and Y[i] ∈ R
The image size used was 158x212 pixels. This is the image size of figure 2.

The following was used to take the facial value which corresponds to figure 2 within the MATLAB program.

```
r=r(55:136,62:124);
g=g(55:136,62:124);
b= b(55:136,62:124);
```

The section to be encrypted was then mapped out as shown in figure 3 and figure 4. The selected section was then encrypted as shown in figure 5 and then placed back as shown in figure 6. An RGB graph of the selected portion of the plain image was then plotted as shown in figure 7 as well as for the encrypted portion as shown in figure 8.

# Conclusion

The pixel displacement and reshuffling of the image in steps between the processes has proven to be really effective. The extra transposition of RGB values in the image file after R G B component reshape has proven the increase of security of the image against all possible attacks available currently. The transposition and displacement technique further makes the operation algorithm to be resistive to li-near deblurring algorithms and other methods of fixing distorted pixels within images.

With this approach it will be difficult for any image deblurring technique using off-the-shelf software to restore the image back to its initial phase. Our future research on this is focused on the employment of public key cryptography in the encryption of images in achieving these goals.

# Biographies

**QUIST-APHETSI KESTER, MIEEE:** is a global award winner 2010 (First place Winner with Gold), in Canada Toronto, of the NSBE's Consulting Design Olympiad Awards and has been recognized as a Global Consulting Design Engineer. Currently the national chair for Policy and Research Internet Society (ISOC) Ghana Chapter, a world renowned body that provides international leadership in Internet related standards, education, and policy. He is the Chairman for the Centre of Research, Information Technology and Advanced computing-CRITAC. He is a law student at the University of London UK. He is a PhD student in Computer Science. The PhD program is in collaboration between the AWBC/ Canada and the Department of Computer Science and Information Technology (DCSIT), University of Cape Coast. He had a Master of Software Engineering degree from the OUM, Malaysia and BSC in Physics from the University of Cape Coast-UCC Ghana.

He has worked in various capacities as a peer reviewer for IEEE ICAST Conference, IET-Software Journal, lecturer, Head of Digital Forensic Laboratory Department at the Ghana Technology University and Head of Computer science department. He is currently a lecturer at the Ghana Technology University College and He may be reached at kquist-aphetsi@gtuc.edu.gh or kquist@ieee.org.